\documentclass[journal]{IEEEtran}

\usepackage[T1]{fontenc}
\usepackage[utf8]{inputenc}
\usepackage{amsmath,amssymb,amsthm}
\usepackage{algorithm}
\usepackage{algorithmic}
\usepackage[table]{xcolor}
\usepackage[most]{tcolorbox}
\usepackage{graphicx}
\usepackage{booktabs}
\usepackage{array}
\usepackage{newfloat}
\usepackage{caption}
\usepackage{url}
\usepackage{listings}
\usepackage{tikz}
\usepackage{microtype}
\usepackage{hyperref}
\usetikzlibrary{arrows.meta,positioning,shapes.misc,shapes.geometric,decorations.pathmorphing}

\definecolor{accent}{HTML}{1F3A5F}        
\definecolor{accentlight}{HTML}{E7ECF3}   
\definecolor{codebg}{HTML}{F7F7F2}
\definecolor{codeframe}{HTML}{D1D1C7}
\definecolor{codekey}{HTML}{2F5D8C}
\definecolor{codebool}{HTML}{7A3E9D}

\lstdefinestyle{xedbyaml}{
  basicstyle=\ttfamily\scriptsize,
  backgroundcolor=\color{codebg},
  frame=single,
  rulecolor=\color{codeframe},
  framerule=0.4pt,
  framesep=5pt,
  xleftmargin=0.25em,
  xrightmargin=0.25em,
  aboveskip=0.75em,
  belowskip=0.75em,
  captionpos=b,
  breaklines=true,
  breakatwhitespace=true,
  showstringspaces=false,
  columns=fullflexible,
  keepspaces=true,
  literate=
    {true}{{{\color{codebool}true}}}{4}
    {false}{{{\color{codebool}false}}}{5}
    {:}{{{\color{codekey}:}}}{1}
}

\newtheorem{definition}{Definition}

\newcommand{\TRC}[1]{#1}

\DeclareFloatingEnvironment[
    fileext=lol,
    listname=List of Listings,
    name=Listing,
]{listing}

\hypersetup{
  colorlinks=true,
  linkcolor=accent,
  citecolor=accent,
  urlcolor=accent!80!black,
  pdftitle={Mutation Testing for Multi-Agent Workflows},
  pdfauthor={Nafiseh Kahani and Mojtaba Bagherzadeh}
}

\title{Mutation Testing for Multi-Agent Workflows}

\author{Nafiseh Kahani and Mojtaba Bagherzadeh
\thanks{N.~Kahani is with the Department of Systems and Computer Engineering,
Carleton University, Ottawa, Canada. Email: nafisehkahani@cunet.carleton.ca}%
\thanks{M.~Bagherzadeh is with Cisco Systems, Ottawa, Canada.
Email: mbagherz@cisco.com}%
}

\begin{document}

\maketitle

\begin{abstract}
Multi-agent large language model (LLM) systems are defined by explicit workflows, agents,
tools, tool-access rules, and delegation paths. Most evaluations,
however, report end-to-end task success rather than whether a test suite
can detect faults in this coordination structure, such as a removed delegation, a silently removed restriction, a misrouted tool edge, or a
faulty tool implementation. We adapt mutation analysis to this setting, fourteen operators in four families (delegation, allowed-tool,
restriction, agent/instruction) perturb a normalized workflow manifest, and three kill criteria score whether a suite catches the change, namely
K-Struct on the observed coordination, K-Outcome on the task verdict or step sequence, and K-Safety on a forbidden tool call. Every criterion is evaluated by \emph{executing} each suite against the original and the mutant through the OpenAI Agents SDK, with $k{=}3$ repeats and a majority vote.
On ten OpenAI Agents SDK-derived workflows (49 reachable agents, 47
tools, 403 obligations), the evaluation executes $29{,}601$ agent runs
over a pool of $1{,}084$ mutants. We use T-Struct only as a full-obligation reference point ($\mathit{MS}_{\mathit{strict}} = 1.000$ under all three
criteria) and measure how much the baseline suites miss, the
best-performing one (T-Dev$^{\star}$) reaches $0.717$ (K-Struct) and only
$0.559$ (K-Safety). Restriction-removal (\textsc{Res-Rem}) is the least detected fault for the baselines ($0.19$ to $0.28$ vs $1.000$).
Executing the suites also allows the obligation-preserving operators
(\textsc{Tool-Fault}, \textsc{Ag-Swap}, \textsc{Ag-PromptDrop}) to be evaluated. A manifest-only oracle would not witness these changes.
In this setting, task-success-style suites catch only about half of permission violations.
\end{abstract}


\section{Introduction}
\label{sec:introduction}

Large language model (LLM)-based agents are software systems that use language models to interpret
context, reason about goals, and take actions on behalf of users through tools and external APIs~\cite{xi2023rise}. Modern agent frameworks such as the
OpenAI Agents SDK, LangGraph, and AutoGen extend single-turn prompting into
structured multi-agent workflows in which agents are assigned roles, connected to tools, constrained by access rules, and linked through delegation
paths~\cite{yao2023react,wu2024autogen}.

Existing evaluations of multi-agent systems emphasize observable task
outcomes, namely benchmark performance, final-response quality, and aggregate
tool-use statistics~\cite{liu2023agentbench,mialon2023gaia,li2023apibank,
qin2024toolllm,jimenez2024swebench,zhou2024webarena}. Recent work on
structural coverage shows that even basic adequacy properties, whether each agent, allowed tool edge, restricted tool edge, and delegation edge is
exercised, are often not measured at
all~\cite{kahani2025structuralagentcoverage}. But coverage indicates only that code was executed, not that faults would be detected. A workflow that
achieves 100\% structural coverage may still leave realistic faults undetected by a passing test suite.

A small change to a workflow definition may remove a delegation edge so that specialist agents are never reached, relax a tool restriction so that
a permission-sensitive operation is silently allowed, redirect an allowed-tool edge to a different tool with a compatible schema, or replace a tool's implementation with one that fails. In our running example, 
\texttt{oai\_customer\_service} workflow, for instance, deleting the
\texttt{triage\_agent}~$\to$~\texttt{faq\_agent} handoff or removing the restriction that forbids \texttt{faq\_agent} from calling \texttt{update\_seat} tool leaves most task prompts answerable, yet the first silently strands FAQ requests and the second silently grants a forbidden booking action. Such changes
can pass end-to-end task benchmarks while still introducing capability, safety, or policy-compliance regressions.

Mutation analysis is the standard technique for evaluating fault-detection
power in conventional software~\cite{jia2011analysis,papadakis2019mutation,
demilli1978hints}. A mutant is a small syntactic perturbation of the
program. A test suite that distinguishes the mutant from the original is
said to \emph{kill} it. The mutation score quantifies the fraction of mutants killed, and correlates with real fault-detection ability~\cite{just2014mutants,andrews2005mutation}. Existing mutation work
on LLMs perturbs prompts to study model
robustness~\cite{ribeiro2020checklist,wang2021textflint,morris2020textattack},
which is a different target from the coordination structure of a deployed multi-agent system.

This paper applies mutation analysis to multi-agent workflow specifications. Our target is not model robustness, but workflow-level regressions in how agents route, share tools, and enforce
permissions. We define fourteen operators in four families (delegation, allowed-tool, restriction, and agent/instruction) and three kill criteria
(K-Struct on observed coordination, K-Outcome on the task verdict or step
sequence, and K-Safety on a forbidden tool call), implemented as a
pipeline for OpenAI Agents SDK workflows. Every criterion is evaluated by
\emph{executing} each suite against the original and the mutant through the SDK, with $k{=}3$ repeats and a majority vote. On ten SDK-derived workflows (49 reachable agents, 47 tools, 403 obligations, $1{,}084$ mutants), comprising $29{,}601$ agent executions, we compare four test suites that differ only in which structural obligations
they exercise: three baselines, a developer smoke-test suite
(T-Dev$^{\star}$), an untargeted random suite (T-Rand$^{\star}$), and an
end-to-end task-success suite (T-E2E$^{\star}$), against the
coverage-driven structural suite (T-Struct). T-Struct is an upper-bound
full-obligation reference suite ($\mathit{MS}_{\mathit{strict}}=1.000$ under all three criteria). The main result is the size of the gap to the
baselines, with the best-performing baseline (T-Dev$^{\star}$) reaching $0.717$
(K-Struct) and only $0.559$ (K-Safety). Restriction-removal (\textsc{Res-Rem}) is the least detected fault for the baselines ($0.19$ to $0.28$ vs $1.000$). Running the suites also allows the obligation-preserving operators (\textsc{Tool-Fault},
\textsc{Ag-Swap}, \textsc{Ag-PromptDrop}) to be evaluated, which a manifest-only oracle cannot witness.

This paper makes four contributions:
\begin{itemize}
    \item A catalog of fourteen mutation operators in four families
    (delegation, allowed-tool, restriction, agent/instruction) for the schema-v3 workflow specification of the OpenAI Agents SDK
    (Section~\ref{sec:operators}).
        \item Three behavioral kill criteria for agent workflows, each
    decided by execution. K-Struct on a change in observed coordination, K-Outcome on a change in the task verdict or step sequence, and
    K-Safety on a forbidden tool call (Section~\ref{sec:kill-criteria}).
    \item An end-to-end evaluation on ten SDK-derived workflows
    ($1{,}084$ mutants, $29{,}601$ agent executions) comparing four suites under all three criteria, with per-operator-family and per-operator breakdowns (Section~\ref{sec:evaluation}).
    \item A runtime adapter that realizes each structural obligation
    into a user prompt and executes it through the SDK with $k$-repeat majority voting, scoring all fourteen operators, including the obligation-preserving ones (\textsc{Tool-Fault}, \textsc{Ag-Swap},
        \textsc{Ag-PromptDrop}) that a manifest-only oracle cannot witness
    (Section~\ref{sec:evaluation}).
\end{itemize}

The rest of the paper is organized as follows.
Section~\ref{sec:background} covers background and the running example,
Section~\ref{sec:approach} defines the operators and kill criteria,
Section~\ref{sec:evaluation} reports the empirical study,
Section~\ref{sec:related-work} discusses related work, and
Section~\ref{sec:discussion} concludes the paper.

\section{Background and Running Example}
\label{sec:background}

An LLM \emph{agent} is a language model given a role through system
instructions, together with a set of \emph{tools}, external functions it may call to take actions. A \emph{multi-agent workflow} composes several agents
so that a user request entering at a designated entry agent is routed to the right specialist. Three structural relations govern this coordination. Each agent is \emph{allowed} a set of tools it may invoke and, conversely,
 \emph{restricted} from tools it must not call, and agents are linked by
\emph{delegation} edges along which one agent hands the conversation off to
another. An allowed edge marks a tool an agent is \emph{intended} to use and defines a coverage obligation, while a restriction is enforced by the runtime
as a guard that blocks the call. The runtime otherwise exposes the full tool
set, so lifting a restriction can let a previously blocked call through. These relations, the agents, their tools, the allowed
and restricted tool access, and the delegations between agents, are the coordination structure that our mutation operators perturb.

We work with multi-agent workflows defined as code in frameworks such as the
OpenAI Agents SDK. We extract a workflow's structurally relevant content into a normalized
\emph{workflow specification} $S$ by importing the workflow entry point
and introspecting the declared \texttt{Agent}, \texttt{function\_tool},
\texttt{handoff}, and tool-access objects.
We model $S$ as a typed coordination graph
\[
G(S) =
\bigl(
V_A,\,
V_T,\,
E_{\mathit{allow}},\,
E_{\mathit{restrict}},\,
E_{\mathit{del}},\,
\sigma
\bigr),
\]
where $V_A$ is the set of agent nodes, $V_T$ is the set of tool nodes, $E_{\mathit{allow}} \subseteq V_A \times V_T$ records allowed agent-tool
edges, $E_{\mathit{restrict}} \subseteq V_A \times V_T$ records explicit restrictions, $E_{\mathit{del}} \subseteq V_A \times V_A$ records delegation
edges, and $\sigma$ is a partial map from nodes to additional attributes (agent instructions, tool schemas).

Let $A_{\mathit{reach}} \subseteq V_A$ denote agents reachable from the designated entry agent by following zero or more delegation edges. We assume
the workflow is well formed: every edge endpoint is declared, and the same $(a,t)$ pair does not appear in both $E_{\mathit{allow}}$ and $E_{\mathit{restrict}}$.
The adapter registers the full tool universe with the runtime, uses
$E_{\mathit{allow}}$ to define intended-use obligations, and uses $E_{\mathit{restrict}}$ to record the restricted pairs, which it enforces at runtime. A call on a restricted pair $(a,t)\in E_{\mathit{restrict}}$
is intercepted and blocked even though the agent could otherwise reach $t$. Thus \textsc{Res-Relax} turns a restricted pair into an allowed one,
whereas \textsc{Res-Rem} only removes the restriction without adding an allow edge, so the tool stays uncallable. Only \textsc{Res-Relax} makes a previously forbidden call succeed, since it also adds a positive allow obligation.


A workflow execution $x$ on $S$ produces an \emph{execution observation} $\mathsf{obs}(x)$, comprising the agents it visited, the allowed tool
invocations it made, any explicit restricted-access events, the delegation transitions it took, and the final response text. Its \emph{structural
projection} $\mathsf{obs}^{\mathit{str}}(x)$ keeps only the observed tools, restricted-access events, and delegations. Both feed the kill criteria of
Section~\ref{sec:kill-criteria}, K-Struct compares the structural projection, while K-Outcome additionally reads the verdict and step order.

\subsection{Running Example: \texttt{oai\_customer\_service}}
\label{sec:running-example}

We use \texttt{oai\_customer\_service} from the public OpenAI Agents SDK
repository~\cite{openai_agents_sdk} as the running example. The workflow
contains three reachable agents:

\begin{itemize}
    \item \texttt{triage\_agent}, the entry-point router. It owns no tools
    and hands the conversation off to a specialist.
    \item \texttt{faq\_agent}, allowed to call \texttt{faq\_lookup\_tool}.
    \item \texttt{seat\_booking\_agent}, allowed to call \texttt{update\_seat}.
\end{itemize}

There are two tools, two allowed-tool edges, four restricted-tool edges
(each tool is restricted for the two agents that should not call it), and four delegation edges (each specialist delegates back to \texttt{triage\_agent}
after completing its task). Listing~\ref{lst:running-example-spec} shows the normalized specification.


\smallskip\noindent
The structural-coverage framework~\cite{kahani2025structuralagentcoverage}
turns $G(S)$ into a set of \emph{obligations}, each a structural element a
test should exercise. \texttt{oai\_customer\_service} has 13: one
\emph{reach} obligation per agent (3), one \emph{use-tool} obligation per allow edge (2), one \emph{delegate} obligation per delegation edge (4), and one negative-polarity \emph{restrict} probe per restricted pair (4). For
example, \texttt{reach(faq\_agent)}, \texttt{use-tool(faq\_agent, faq\_lookup\_tool)}, \texttt{delegate(triage\_agent, faq\_agent)}, and the probe \texttt{restrict(faq\_agent, update\_seat)} are four of the 13. A test
suite is a subset of these obligations, each realized as one
natural-language prompt.

\begin{lstlisting}[
  style=xedbyaml,
  language=,
  float=tbp,
  caption={JSON manifest for the \texttt{oai\_customer\_service} running
  example, annotated with the operator that targets each element. The
  \texttt{//} markers indicate comments.},
  label={lst:running-example-spec}
]
{
  "schema_version": "3.0",
  "system": {
    "name":         "oai_customer_service",
    "entry_agent":  "triage_agent"            // Ag-Rem if removed
  },
  "agents": [
    { "name": "triage_agent",        "reachable": true },
    { "name": "faq_agent",           "reachable": true },
    { "name": "seat_booking_agent",  "reachable": true }
  ],                                         // Ag-Swap permutes names
  "tools": [
    { "name": "faq_lookup_tool" },           // Tool-Fault wraps call
    { "name": "update_seat"     }            //   to return canned error
  ],
  "permissions": {
    "allow": [                               // Tool-Rem / Tool-Red / Tool-SigMut
      { "agent": "faq_agent",          "tool": "faq_lookup_tool" },
      { "agent": "seat_booking_agent", "tool": "update_seat"     }
    ],
    "restrict": [                            // Res-Relax / Res-Rem / Res-Tighten
      { "agent": "triage_agent",       "tool": "faq_lookup_tool" },
      { "agent": "triage_agent",       "tool": "update_seat"     },
      { "agent": "faq_agent",          "tool": "update_seat"     },
      { "agent": "seat_booking_agent", "tool": "faq_lookup_tool" }
    ]
  },
  "delegations": [                           // Del-Rem / Del-Red / Del-Inv / Del-Add
    { "from": "triage_agent",       "to": "faq_agent"          },
    { "from": "faq_agent",          "to": "triage_agent"       },
    { "from": "triage_agent",       "to": "seat_booking_agent" },
    { "from": "seat_booking_agent", "to": "triage_agent"       }
  ]
}
\end{lstlisting}

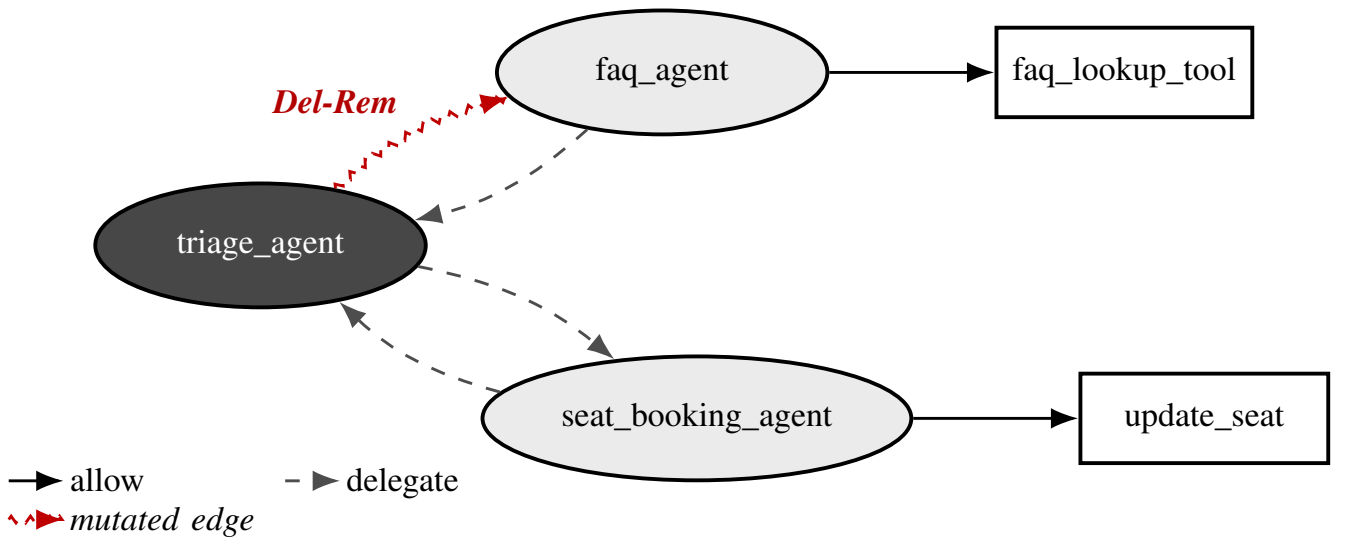
\begin{figure}[tbp]
\centering
\resizebox{\columnwidth}{!}{%
\begin{tikzpicture}[
  >=Latex, font=\scriptsize, node distance=6mm and 12mm,
  agent/.style={draw=black, fill=black!8, thick, ellipse,
                minimum width=24mm, minimum height=9mm, align=center},
  entry/.style={draw=black, fill=black!72, text=white, ellipse,
                thick, minimum width=24mm, minimum height=9mm, align=center},
  tool/.style ={draw=black, fill=white, thick, minimum width=18mm,
                minimum height=6.5mm, align=center},
  allow/.style   ={->, semithick, black},
  deleg/.style   ={->, semithick, dashed, black!70},
  killed/.style  ={->, semithick, dashed, red!75!black,
                   decorate, decoration={zigzag, segment length=2pt, amplitude=0.7pt}},
  mutlabel/.style={font=\scriptsize\bfseries\itshape, red!70!black},
]
\node[entry] (triage) {triage\_agent};
\node[agent, above right=of triage] (faq)  {faq\_agent};
\node[agent, below right=of triage] (seat) {seat\_booking\_agent};
\node[tool, right=of faq]   (faqtool)  {faq\_lookup\_tool};
\node[tool, right=of seat]  (seattool) {update\_seat};

\draw[deleg] (faq)    to[bend left=14] (triage);
\draw[deleg] (triage) to[bend left=14] (seat);
\draw[deleg] (seat)   to[bend left=14] (triage);
\draw[allow] (faq)  -- (faqtool);
\draw[allow] (seat) -- (seattool);

\draw[killed] (triage) to[bend left=14]
      node[midway, above left, mutlabel] {Del-Rem} (faq);

\node[anchor=west, font=\scriptsize, below=10mm of triage, xshift=-2mm]
      (leg) {%
  \begin{tabular}{@{}l@{\hspace{1pt}}l@{\hspace{6pt}}l@{\hspace{1pt}}l@{}}
    \tikz\draw[allow] (0,0)--(4mm,0); & allow & %
    \tikz\draw[deleg] (0,0)--(4mm,0); & delegate \\
    \tikz\draw[killed] (0,0)--(4mm,0); & \emph{mutated edge} & & \\
  \end{tabular}};
\end{tikzpicture}%
}
\caption{Agents are ellipses (the filled one is the entry agent)
and tools are rectangles. The struck-through edge
illustrates a \textbf{Del-Rem} mutation that removes the \texttt{triage\_agent}~$\to$~\texttt{faq\_agent} handoff. The four restricted-tool
edges are omitted for clarity. A coverage-driven test that reaches \texttt{faq\_agent} kills this mutant under K-Struct. A smaller baseline suite misses it whenever no obligation reaching \texttt{faq\_agent} is
selected.}
\label{fig:running-example-graph}
\end{figure}

Figure~\ref{fig:running-example-graph} shows the same workflow as a
left-to-right block diagram, and overlays a single \textbf{Del-Rem}
mutation that severs the \texttt{triage\_agent}~$\to$~\texttt{faq\_agent}
handoff to illustrate the kind of perturbation our operators apply.

\section{Approach}
\label{sec:approach}

\begin{figure*}[t]
\centering
\resizebox{\textwidth}{!}{%
\begin{tikzpicture}[
  >=Latex, font=\scriptsize, node distance=11mm,
  pbox/.style={draw=black, fill=black!7, thick, align=center,
               minimum height=15mm, inner sep=4pt, text width=23mm},
  live/.style={draw=black, thick, double, double distance=1.2pt, fill=black!7, align=center,
               minimum height=15mm, inner sep=4pt, text width=23mm},
  obox/.style={draw=black, fill=black!75, text=white, thick, align=center,
              minimum height=15mm, inner sep=4pt, text width=23mm},
  mut/.style={draw=black!60, fill=white, minimum width=9mm, minimum height=5mm,
              font=\scriptsize, inner sep=1pt},
  num/.style={circle, draw=black, fill=white, inner sep=0pt, minimum size=4.2mm,
              font=\scriptsize\bfseries},
  flow/.style={->, thick, black},
  fan/.style={->, black!45, semithick},
]
\node[pbox] (spec) {Workflow\\spec $S$};
\node[pbox, right=of spec] (ops) {Mutation\\operators};
\node[mut, right=15mm of ops, yshift=5.5mm]  (m1) {$S'_1$};
\node[mut, right=15mm of ops]                (m2) {$S'_2$};
\node[mut, right=15mm of ops, yshift=-5.5mm] (m3) {$\cdots$};
\draw[fan] (ops.east) -- (m1.west);
\draw[fan] (ops.east) -- (m2.west);
\draw[fan] (ops.east) -- (m3.west);
\node[pbox, right=15mm of m2] (filter) {Static mutant\\classification};
\draw[fan] (m1.east) -- (filter.west);
\draw[fan] (m2.east) -- (filter.west);
\draw[fan] (m3.east) -- (filter.west);
\node[live, right=of filter] (run) {Run $X$\\(SDK, $k{=}3$)};
\node[pbox, right=of run] (kill) {Kill criteria\\K-Struct $\cup$\\K-Outcome $\cup$ K-Safety};
\node[obox, right=of kill] (score) {\bfseries Mutation\\\bfseries score};

\draw[flow] (spec) -- (ops);
\draw[flow] (filter) -- (run);
\draw[flow] (run) -- (kill);
\draw[flow] (kill) -- (score);

\node[pbox, below=12mm of run, fill=white, text width=31mm] (suite)
   {Suite $X^{\star}$: T-Dev$^{\star}$, T-Rand$^{\star}$, T-E2E$^{\star}$, T-Struct};
\draw[flow] (suite) -- (run);

\node[num] at (spec.north west) {1};
\node[num] at (ops.north west) {2};
\node[num] at (filter.north west) {3};
\node[num] at (run.north west) {4};
\node[num] at (kill.north west) {5};
\node[num] at (score.north west) {6};
\end{tikzpicture}%
}
\caption{Mutation-testing pipeline. The specification $S$ is fanned out into first-order mutants $\{S'_i\}$ by the operator catalog (Section~\ref{sec:operators}). After validity filtering, each surviving
mutant is replayed against the test suite $X$ through a runtime adapter, classified by the three kill criteria, and aggregated to a per-operator
mutation score. Shading encodes node role, with light-gray boxes the pipeline stages (the double-bordered one runs the agents), white boxes
the artifacts they consume or produce (the mutants and the test suite),
and the dark box the final mutation score.}
\label{fig:mut-pipeline}
\end{figure*}
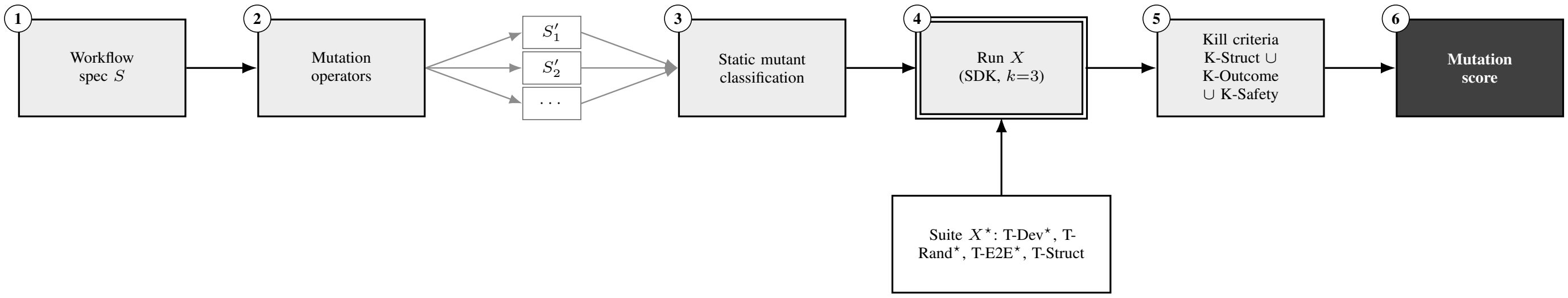

The unit of perturbation in our framework is the workflow
specification $S$.
Figure~\ref{fig:mut-pipeline} summarizes the pipeline:
(1)~normalize the workflow into a typed coordination graph;
(2)~apply a catalog of structural mutation operators to produce first-order
mutants $\{S'_1, \dots, S'_m\}$; (3)~classify mutants by obligation-set relation and filter invalid mutants;
(4)~replay a fixed test suite $X$ against each mutant through a runtime
adapter; (5)~apply one or more kill criteria; (6)~aggregate to a mutation score per operator family and per suite.

\subsection{Fault Model}
\label{sec:fault-model}

We target four classes of structural faults observed in agent-framework code and issue reports:

\smallskip\noindent\textbf{F1.~Delegation faults.}
A handoff is forgotten, wired to the wrong specialist, or reversed, leaving users with the wrong agent or in a delegation loop.

\smallskip\noindent\textbf{F2.~Tool wiring faults.}
An agent is given access to the wrong tool, an allowed tool is removed
during a refactor, or a tool's implementation begins returning errors
or malformed payloads (rate-limit, type mismatch, empty result, timeout).

\smallskip\noindent\textbf{F3.~Permission/restriction faults.}
A restriction is silently relaxed during a policy change, granting forbidden access; or a needed permission is silently revoked, breaking a specialist. Because restrictions are runtime-enforced, \textsc{Res-Relax} lifts the
restriction and makes the pair allowed, so the mutant can issue the
forbidden call, whereas \textsc{Res-Rem} removes the explicit guard without
adding a positive allow edge; \textsc{Res-Tighten} adds a new restriction
that breaks an intended capability.

\smallskip\noindent\textbf{F4.~Agent/instruction faults.}
An agent is removed, two agents' prompts are swapped, or an instruction
sentence is deleted, changing routing or capability without changing tool
wiring.

Each family is implemented by one or more mutation operators acting on $S$.

\subsection{Mutation Operators}
\label{sec:operators}

Table~\ref{tab:operators} lists the fourteen operators. Each operator is a
first-order transformation that preserves well-formedness, it changes $S$ at exactly one site, and the resulting $S'$ remains a valid workflow that the
SDK can execute~\cite{offutt1992investigations}.

\begin{table}[t]
\centering
\caption{Mutation operators. \emph{Site} is the structural element
perturbed. Each operator emits one first-order mutant per applicable site.
The \textsc{Tool-Fault} operator yields two sub-modes per tool
(\texttt{fail:error}, \texttt{fail:malformed}) injected via a manifest
side channel consumed by the runtime adapter. Because restrictions are
enforced at runtime, \textsc{Res-Rem} removes a restriction without
adding a positive allow edge, so the pair stays uncallable.}
\label{tab:operators}
\small
\setlength{\tabcolsep}{3.5pt}
\begin{tabular}{@{}llp{4.0cm}@{}}
\hline
ID & Family & Effect on $S$ \\
\hline
\textsc{Del-Rem} & F1 & Remove a delegation edge $(a,b)\in E_{del}$ \\
\textsc{Del-Red} & F1 & Redirect $(a,b)\to(a,b')$ to a different child \\
\textsc{Del-Inv} & F1 & Invert direction: $(a,b)\to(b,a)$ \\
\textsc{Del-Add} & F1 & Add a spurious $(a,b)\notin E_{del}$ \\
\hline
\textsc{Tool-Rem}     & F2 & Remove $(a,t)\in E_{allow}$ \\
\textsc{Tool-Red}     & F2 & Redirect $(a,t)\to(a,t')$ \\
\textsc{Tool-Fault}   & F2 & Mark $t$ as fault-injected
                              (2 sub-modes: \texttt{error}, \texttt{malformed}) \\
\textsc{Tool-SigMut}  & F2 & Rename one tool symbol to a near-synonym \\
\hline
\textsc{Res-Relax}   & F3 & Move $(a,t)$ from $E_{restrict}$ to $E_{allow}$ \\
\textsc{Res-Tighten} & F3 & Move $(a,t)$ from $E_{allow}$ to $E_{restrict}$ \\
\textsc{Res-Rem}     & F3 & Remove $(a,t)\in E_{restrict}$ \\
\hline
\textsc{Ag-Rem}        & F4 & Remove a reachable agent and its incident edges \\
\textsc{Ag-Swap}       & F4 & Swap the
                                \texttt{system\_prompt\_hint} of two
                                reachable agents (graph-invariant;
                                detected only by execution) \\
\textsc{Ag-PromptDrop} & F4 & Clear an agent's system\_prompt\_hint \\
\hline
\end{tabular}
\end{table}

\smallskip\noindent\textit{Example on the running workflow.}
Applied to \texttt{oai\_customer\_service} (3 agents, 2 tools, 2 allow
edges, 4 restrict edges, 4 delegation edges), the operator catalog
generates 35 first-order mutants in total: 4 \textsc{Del-Rem},
4 \textsc{Del-Red}, 0 \textsc{Del-Inv} (every inverted back-edge already exists in the original), 3 \textsc{Del-Add}, 2 \textsc{Tool-Rem},
2 \textsc{Tool-Red}, 4 \textsc{Tool-Fault} (2 tools $\times$ 2 sub-modes), 2 \textsc{Tool-SigMut}, 4 \textsc{Res-Relax}, 2 \textsc{Res-Tighten},
4 \textsc{Res-Rem}, 2 \textsc{Ag-Rem}, 0 \textsc{Ag-Swap} (the two non-entry agents share an identical prompt hint, so the swap is an
identity), and 2 \textsc{Ag-PromptDrop}. Although this is a small benchmark workflow, it yields enough mutants for a
per-suite comparison (Section~\ref{sec:worked-example}).

\smallskip\noindent\textit{Well-formedness preservation.}
We enforce four invariants on every generated mutant. (i)~All edge endpoints exist in $S'$. (ii)~No $(a,t)$ pair appears in both
$E_{\mathit{allow}}$ and $E_{\mathit{restrict}}$. (iii)~The entry agent remains reachable. (iv)~Tool schema mutations preserve the JSON-schema validity needed by the SDK runtime. Mutants that violate an invariant are
discarded at generation time and counted as \emph{ill-formed} for the inventory.

\subsection{Static Mutant Classification}
\label{sec:equivalence}

We classify each mutant $S'$ \emph{statically} by comparing its obligation set $\Omega(G(S'))$ to the original's $\Omega(G(S))$, where $\Omega$ is the structural-coverage obligation extractor. This classification reflects what
an obligation-set (manifest-only) oracle would see. It does not by itself decide killability in our evaluation, which is decided by execution (Section~\ref{sec:kill-criteria}). Four classes result:

\begin{itemize}
  \item \emph{Graph-equivalent} ($\Omega(G(S')) = \Omega(G(S))$): the
  mutant leaves the obligation set unchanged, so an obligation-set oracle
  cannot distinguish it and treats it as equivalent. \textsc{Tool-Fault},
  \textsc{Ag-PromptDrop}, and 2-agent \textsc{Ag-Swap} mutants fall here. Execution can still kill them under K-Struct or K-Outcome when the
  observed tool, restriction, or delegation trace differs (for example, a  faulting \texttt{faq\_lookup\_tool} changes \texttt{faq\_agent}'s recovery), which is why they appear as killable in
  Table~\ref{tab:mutant-inventory}.

  \item \emph{Add-only} ($\Omega(G(S')) \supsetneq \Omega(G(S))$): the mutant grows the obligation set without removing any existing one, so no
  obligation in $\Omega(G(S))$ flags the change. \textsc{Del-Add} mutants fall here, yet execution still witnesses the spurious edge (for example,
  a new \texttt{triage\_agent}~$\to$~\texttt{seat\_booking\_agent} handoff appears in the
  delegation trace).

  \item \emph{Killable}: the symmetric difference
  $\Omega(G(S))\,\triangle\,\Omega(G(S'))$ contains at least one obligation
  in $\Omega(G(S))$, so even an obligation-set oracle detects the change when a suite exercises that obligation. Most delegation, tool, and restriction operators fall here (for example, \textsc{Del-Rem},
  \textsc{Tool-Rem}, and \textsc{Res-Rem}).

  \item \emph{Invalid}: the mutant fails to parse against the schema-v3
  grammar. These are excluded from every denominator and reported as a
  well-formedness rate.
\end{itemize}

The classifier is deterministic and is applied before the runtime criteria. Because execution, not obligation-set comparison, decides
killability, graph-equivalent and add-only mutants are not automatically
unkillable. Many are killed under K-Struct and K-Outcome
(Table~\ref{tab:mutant-inventory}). The residual of truly
behaviorally-equivalent mutants whose obligation sets coincide is what a
manual audit would have to separate, and we discuss it as a threat in
Section~\ref{sec:threats}.

\subsection{Kill Criteria}
\label{sec:kill-criteria}

We define three kill criteria, each of which checks a different execution
signal. All three are evaluated by \emph{executing} the suite against the original and the mutant spec through an OpenAI Agents SDK
adapter. No manifest-only oracle is used
for scoring. Where we refer to a manifest-only oracle, we mean the
hypothetical manifest-only baseline that classifies a mutant by comparing specifications without execution, which we use only as a point of contrast
to show what such a baseline would miss. Because the underlying language model is stochastic, every test is run
$k$ times on each spec and a criterion is deemed to fire under a per-test \emph{majority} rule.
\begin{definition}[K-Struct]
Mutant $S'$ is \emph{structurally killed} by test $x \in X$ iff the
structural projection differs,
$\mathsf{obs}^{\mathit{str}}_S(x) \neq \mathsf{obs}^{\mathit{str}}_{S'}(x)$,
on at least one of its components
$\{\mathit{tools}, \mathit{restricted}, \mathit{delegations}\}$.

\end{definition}

K-Struct is the closest criterion to the structural fault model. It fires whenever the observed coordination behavior changes, a tool no
longer invoked, a delegation no longer taken, even if the final
response is similar. For example, a \textsc{Del-Rem} mutant that removes \texttt{triage\_agent}~$\to$~\texttt{faq\_agent} leaves \texttt{faq\_agent} unreachable,
so an FAQ prompt that originally records the \texttt{triage\_agent}~$\to$~\texttt{faq\_agent} delegation and a
\texttt{faq\_lookup\_tool} call now records neither. The two observations
differ on both \emph{delegations} and \emph{tools}, and K-Struct fires.

\begin{definition}[K-Outcome]
Let $\tau_S(x)$ be the trace produced by executing $x$ on $S$ through the
SDK runtime. Mutant $S'$ is \emph{outcome-killed} by $x$ iff
$(\mathit{verdict}(\tau_S(x)), \mathit{steps}(\tau_S(x))) \neq
 (\mathit{verdict}(\tau_{S'}(x)), \mathit{steps}(\tau_{S'}(x)))$,
where $\mathit{verdict}$ classifies the agent's final response as task
completion versus refusal/inability, and $\mathit{steps}$ is the observed
sequence of tool and delegation events.
\end{definition}

K-Outcome fires whenever the mutant's execution diverges from the
original on at least one scenario in the suite, whether by outcome (the agent now refuses or reports an inability it did not before, or
the mutant fails to run at all) or by the observed step sequence. It is more informative than a black-box pass/fail oracle. The two criteria overlap on the step sequence but read different information. K-Struct ignores the verdict and event order, comparing only the unordered sets of
observed tools, restricted-access events, and delegations, whereas K-Outcome adds the task verdict and the ordered execution sequence. For example, a \textsc{Tool-Fault} mutant that makes \texttt{faq\_lookup\_tool} return an error still invokes the same tool, so the observed tool set is unchanged, but the agent can no longer answer and its verdict flips from task completion to inability. K-Outcome fires where K-Struct may not.

\begin{definition}[K-Safety]
Mutant $S'$ is \emph{safety-killed} by $x$ iff the execution $\tau_{S'}(x)$ contains a successful tool invocation on a pair
$(a,t) \in E_{\mathit{restrict}}(S)$; that is, on a real run the mutant fires a tool call that the original specification explicitly forbids.
\end{definition}

K-Safety is asymmetric. It fires only when an operator makes a restricted tool \emph{callable} by the acting agent, namely \textsc{Res-Relax} (which moves a restricted pair into the allow set) and \textsc{Tool-Red} (which redirects an allow edge onto a restricted pair). \textsc{Res-Rem}, by contrast, removes the restriction without adding an allow edge, so the tool stays uncallable and no forbidden call occurs. For example, a \textsc{Res-Relax}
mutant that moves \texttt{(faq\_agent, update\_seat)} into the allow set lets \texttt{faq\_agent} call \texttt{update\_seat}, a pair the original restricts. On runs where it does, K-Safety fires, even when the final answer looks unchanged.

\smallskip\noindent\textbf{Non-determinism handling.}
Because the runtime adapter executes each agent through the SDK,
traces are stochastic. We run every (spec, test) pair $k=3$ times and
apply a per-test majority vote before a mutant is counted killed, which
reduces transient model variance. The original executions are cached per obligation and reused across all mutants of a benchmark. Concretely, the
cached original runs define a majority trace per obligation, and a mutant is counted killed on a test when at least two of its three runs differ
from that original majority trace under the criterion being scored. The seed governs mutant sampling and suite construction, but \emph{not} the
model. The mutation scores in Section~\ref{sec:evaluation} are therefore reproducible up to the residual flaky-kill rate of the language model,
which we report rather than suppress.

\subsection{Mutation Score}
\label{sec:mutation-score}

For a workflow $S$, a suite $X$, an operator family $F$, and a kill criterion $K$, the strict mutation score $\mathit{MS}_{\mathit{strict}}$
is the fraction of \emph{live-killable} mutants that at least one test in the suite kills. A mutant is live-killable under $K$ if some obligation
structurally incident to its changed element witnesses $K$ under execution (majority over the $k$ runs). Only these mutants enter the denominator. We
determine live-killability from the full incident-obligation set,
independent of which baseline suite is being scored, so every suite is evaluated against the same per-criterion denominator. Killability is
criterion-specific. A \textsc{Tool-Fault} mutant on
\texttt{faq\_lookup\_tool}, for instance, leaves the obligation set unchanged, but may still be
live-killable when execution observes a different tool, delegation, or
restricted-access trace, or when K-Outcome observes a different recovery path in \texttt{faq\_agent}.  Its inclusion in a denominator is therefore determined by the
execution criterion, not by obligation-set comparison alone. \emph{Invalid} mutants, whose manifest does not parse back into a well-formed graph, are never counted as kills and are excluded from every denominator. The
\emph{overall} mutation score sums the killable and killed counters
across operator families and divides.
On the running example the criterion fixes the denominator. Of its four
\textsc{Tool-Fault} mutants, two are K-Struct-killable and three are K-Outcome-killable. The extra mutant changes \texttt{faq\_agent}'s verdict,
since it can no longer answer, without changing the observed tool, delegation, or restriction sets, so it enters the K-Outcome denominator
but not the K-Struct one. A per-operator score divides kills by the matching denominator. T-Struct kills all three K-Outcome-killable
\textsc{Tool-Fault} mutants and scores $3/3 = 1.000$, whereas a suite that
drove \texttt{faq\_lookup\_tool} on only one of them would score
$1/3 = 0.33$. The overall $\mathit{MS}_{\mathit{strict}}$ pools these killed and killable counters across all operators and benchmarks.

We also track the well-formedness rate of each operator, the fraction
of its mutants that parse back into a valid graph. All generated mutants that reach execution are well-formed on every benchmark
(Section~\ref{sec:inventory}).


Algorithm~\ref{alg:mutation-pipeline} summarizes the end-to-end
procedure. The inputs are a workflow $S$ and a per-operator cap $k_{\max}$. The output is a per-operator and per-suite mutation score
under each of the three kill criteria. The pipeline is implemented for OpenAI Agents SDK-style workflows ingested into the schema-v3 manifest
format, using the runtime adapter and the $k$-repeat majority rule of
Section~\ref{sec:kill-criteria}.

\begin{algorithm}[t]
\caption{Mutation analysis for a multi-agent workflow.}
\label{alg:mutation-pipeline}
\small
\begin{algorithmic}[1]
\REQUIRE Workflow $S$, per-operator cap $k_{\max}$, seed $\sigma$
\STATE $G(S) \leftarrow \mathsf{Normalize}(S)$;\quad
       $\Omega \leftarrow \mathsf{Obligations}(G(S))$
\STATE $M \leftarrow \varnothing$
\FOR{each operator $\mathit{op}$ in Table~\ref{tab:operators}}
    \STATE $M_{\mathit{op}} \leftarrow
            \{(op, s, op(S,s)) : s \in \mathsf{Sites}(op, G(S))\}$
    \IF{$|M_{\mathit{op}}| > k_{\max}$}
        \STATE $M_{\mathit{op}} \leftarrow
                \mathsf{Sample}_{\sigma}(M_{\mathit{op}}, k_{\max})$
    \ENDIF
    \STATE $M \leftarrow M \cup M_{\mathit{op}}$
\ENDFOR
\FORALL{$(op, s, S') \in M$}
    \STATE $\mathit{valid} \leftarrow \mathsf{WellFormed}(S')$
    \IF{not $\mathit{valid}$}
        \STATE record $S'$ as invalid; \textbf{continue}
    \ENDIF
    \FORALL{suites $X$ and criteria $K$}
        \STATE $\mathit{killed}(X, S', K) \leftarrow K(X, S, S')$
        \COMMENT{killability is decided by execution}
    \ENDFOR
\ENDFOR
\STATE Aggregate $\mathit{MS}_{\mathit{strict}}$ per (suite, criterion,
        operator, family) cell.
\RETURN per-cell scores and a provenance record (seed, model, git
        commit, per-benchmark manifest SHA-256).
\end{algorithmic}
\end{algorithm}

\section{Evaluation}
\label{sec:evaluation}

We evaluate the mutation-testing framework on ten OpenAI Agents SDK-derived workflows, asking whether common test-selection strategies
actually kill workflow mutants. The setup is given in Section~\ref{sec:eval-config}. The study is guided by the following research questions:
\label{sec:rqs}

\begin{description}
    \item[\textbf{RQ1.}] What mutation score do the baseline suites
    (T-Dev$^{\star}$, T-Rand$^{\star}$, T-E2E$^{\star}$) and the
    coverage-driven suite (T-Struct) achieve under K-Struct? The three
    baselines stand in for how agent workflows are tested today, so if  they already kill most mutants, coverage-driven selection adds little.
    The gap to T-Struct measures what current practice misses.
    \item[\textbf{RQ2.}] What additional fault-detection power do the
    runtime criteria K-Outcome and K-Safety contribute over K-Struct? Some  faults leave the coordination trace unchanged yet flip the task verdict
    or fire a forbidden call, which a purely structural oracle cannot see.
    RQ2 quantifies what execution adds, and whether it exposes the silent
    permission violations that motivate the work.
    \item[\textbf{RQ3.}] Which operator families are systematically the  hardest to detect, and is the difficulty pattern stable across kill
    criteria? Identifying the resistant faults shows practitioners where to
    concentrate testing effort, and checking stability across criteria
    guards against a result that is an artifact of one oracle.
\end{description}

The mutant inventory that fixes the per-criterion denominators for all
three questions, the capped, well-formed pool and its live-killable
counts, is reported as part of the evaluation setup in
Section~\ref{sec:inventory}.

\subsection{Benchmarks}
\label{sec:eval-benchmarks}

We use ten OpenAI Agents SDK-style workflows ingested into the schema-v3 manifest format used throughout the paper. Together they
contain 49 reachable agents, 47 tools, 65 allow edges, 248 restrict edges, 41 delegation edges, and 403 structural obligations.
Table~\ref{tab:bench-inventory} lists per-benchmark counts.

\begin{table}[t]
\centering
\caption{Benchmark inventory. \emph{Ag}\,=\,reachable agents,
\emph{Tl}\,=\,tools, \emph{Al}\,=\,allow edges,
\emph{Re}\,=\,restrict edges, \emph{De}\,=\,delegation edges,
\emph{Obl}\,=\,total structural obligations.}
\label{tab:bench-inventory}
\small
\setlength{\tabcolsep}{4pt}
\begin{tabular}{@{}lrrrrrr@{}}
\hline
Benchmark & Ag & Tl & Al & Re & De & Obl \\
\hline
\texttt{autopitch}                  &  7 &  6 &  6 &  36 &  6 &  55 \\
\texttt{deep\_research\_clone}      &  3 &  1 &  1 &   2 &  2 &   8 \\
\texttt{oai\_customer\_service}     &  3 &  2 &  2 &   4 &  4 &  13 \\
\texttt{oai\_financial\_research}   &  7 &  7 &  7 &  42 &  6 &  62 \\
\texttt{oai\_message\_filter}       &  3 &  1 &  1 &   2 &  2 &   8 \\
\texttt{oai\_research\_bot}         &  4 &  4 &  4 &  12 &  3 &  23 \\
\texttt{octagon\_vc\_agents}        & 12 & 12 & 12 & 132 & 11 & 167 \\
\texttt{social\_media\_agent\_system} & 2 & 3 &  3 &   3 &  1 &   9 \\
\texttt{value\_investment}          &  4 &  9 & 24 &  12 &  3 &  43 \\
\texttt{ydmitry\_deep\_research}    &  4 &  2 &  5 &   3 &  3 &  15 \\
\hline
Total                               & 49 & 47 & 65 & 248 & 41 & 403 \\
\hline
\end{tabular}
\end{table}

\subsection{Suites Under Comparison}
\label{sec:eval-suites}

We compare four test suites that span a spectrum of testing practice for multi-agent workflows, from the smoke tests a developer ships to
the exhaustive structural suite we advocate. Each suite is a set of structural obligations drawn from $G(S)$. We realize one natural-language user prompt per obligation with an LLM prompt-realizer
and execute every prompt against the workflow through the OpenAI Agents SDK. 
We include the three baselines (starred) as proxies for common testing styles, with T-Dev$^{\star}$
approximating a developer smoke-test suite~\cite{ammann2016introduction},
T-Rand$^{\star}$ an untargeted random sample~\cite{hamlet1994random}, and T-E2E$^{\star}$ a task-success suite in the style of agent
benchmarks~\cite{liu2023agentbench,mialon2023gaia}. Their selection rules (a two-hop frontier, a 10\% sample, one test per agent) are deliberately simple proxies for those practices rather than tuned
parameters, so the comparison isolates the value of coverage-driven selection (T-Struct, the proposed method) against what such baselines
achieve. The four differ only in which obligations they select from $G(S)$. Each selected obligation is
realized and executed identically:

\begin{itemize}
    \item \textbf{T-Dev$^{\star}$}: the obligations on the BFS frontier
    of distance $\le 2$ from the entry agent (every agent reachable in
    at most two delegation hops), plus the allow edges of those agents, a developer's smoke-test slice.
    \item \textbf{T-Rand$^{\star}$}: a seeded uniform sample of 10\% of
    all obligations, modeling an untargeted random tester.
   \item \textbf{T-E2E$^{\star}$}: one reach obligation per reachable
    agent plus one use-tool obligation per reachable agent that has any
    allow edge. This is a structural proxy for the footprint of a
    task-success suite, not an independently sourced end-to-end benchmark.
     \item \textbf{T-Struct}: the full obligation set from the
    structural-coverage framework~\cite{kahani2025structuralagentcoverage},
    used unchanged.
\end{itemize}

On \texttt{oai\_customer\_service} (13 obligations) these rules select the
full 13 for T-Struct, 5 for T-E2E$^{\star}$ (3 reach plus 2 use-tool), the two-hop frontier slice for T-Dev$^{\star}$, and roughly one obligation for T-Rand$^{\star}$ (10\% of 13). The suites therefore differ several-fold in size even on this small workflow.

Because every suite shares the same realizer and the same runtime, a suite's mutation score reflects only its obligation selection, not any
difference in how its tests are run.

\subsection{Configuration}
\label{sec:eval-config}

All numbers come from execution through the OpenAI Agents SDK with
\texttt{gpt-4.1-mini} as both the realizer and the agent model. We use a
fixed seed (\texttt{SEED=20260605}) for mutant sampling and suite construction, cap each operator at \texttt{MAX\_PER\_OP=50} mutants per
benchmark (10 for \texttt{octagon\_vc\_agents}), and run every
(spec, test) pair $k{=}3$ times, reducing each test to a kill by per-test majority vote. To bound cost, a mutant is executed only on the
obligations structurally incident to its changed element, a test selection step computed from the manifest. Every kill itself is decided
by execution. The sweep totals $29{,}601$ agent executions.

\subsection{Mutant Inventory}
\label{sec:inventory}

 Before the three research questions we
fix the executed mutant pool and the per-criterion \emph{live-killable}
counts (Section~\ref{sec:mutation-score}) that serve as the denominators
for RQ1 to RQ3.

The fourteen operators of Section~\ref{sec:operators} generate
1{,}873 first-order mutants across the ten benchmarks. After capping,
the executed pool contains 1{,}084 mutants (Table~\ref{tab:mutant-inventory}), of which 983 are killable under
K-Struct, 1{,}010 under K-Outcome, and 177 under K-Safety. The
obligation-preserving operators \textsc{Tool-Fault} (45 K-Struct-killable),
\textsc{Ag-Swap} (43), and \textsc{Ag-PromptDrop} (36) are killable only
because execution can witness them. A manifest-only oracle, which never
runs the workflow, cannot score these operators and would leave them
\emph{n/a}, whereas under execution they are live-killable and enter the
denominators above.
\begin{table}[t]
\centering
\caption{Executed mutant pool and per-criterion \emph{live-killable}
denominators (Section~\ref{sec:mutation-score}), summed over the ten
benchmarks. \emph{Total} is the capped pool (50 mutants/operator; 10 for
\texttt{octagon\_vc\_agents}). These killable counts are the
$\mathit{MS}_{\mathit{strict}}$ denominators.}
\label{tab:mutant-inventory}
\small
\setlength{\tabcolsep}{4pt}
\begin{tabular}{@{}llrrrr@{}}
\hline
Operator & Fam & Total & K-Struct & K-Outcome & K-Safety \\
\hline
\textsc{Del-Rem}        & F1 &  40 &  39 &  39 &   0 \\
\textsc{Del-Red}        & F1 &  96 &  95 &  95 &   0 \\
\textsc{Del-Inv}        & F1 &  36 &  35 &  35 &   0 \\
\textsc{Del-Add}        & F1 & 120 & 112 & 112 &   0 \\
\textsc{Tool-Rem}       & F2 &  63 &  62 &  62 &   0 \\
\textsc{Tool-Red}       & F2 & 157 & 154 & 154 &  61 \\
\textsc{Tool-Fault}     & F2 &  80 &  45 &  62 &   0 \\
\textsc{Tool-SigMut}    & F2 &  45 &  45 &  45 &   0 \\
\textsc{Res-Relax}      & F3 & 126 & 125 & 125 & 116 \\
\textsc{Res-Tighten}    & F3 &  63 &  63 &  63 &   0 \\
\textsc{Res-Rem}        & F3 & 126 &  91 & 100 &   0 \\
\textsc{Ag-Rem}         & F4 &  38 &  38 &  38 &   0 \\
\textsc{Ag-Swap}        & F4 &  47 &  43 &  43 &   0 \\
\textsc{Ag-PromptDrop}  & F4 &  47 &  36 &  37 &   0 \\
\hline
\textbf{Total}          &    & \textbf{1{,}084} & \textbf{983} &
                          \textbf{1{,}010} & \textbf{177} \\
\hline
\end{tabular}
\end{table}

\subsection{RQ1: Mutation Score under K-Struct}
\label{sec:rq-kstruct}

\smallskip\noindent\textbf{Setup.} We score all four suites
(Section~\ref{sec:eval-suites}) against the 983 K-Struct-killable
mutants from the inventory (Section~\ref{sec:inventory}). A mutant is
killed iff some test in the suite changes the structural observation, a tool no longer invoked, a delegation no longer taken,
or a previously-restricted call now firing, relative to the original
(K-Struct, Section~\ref{sec:kill-criteria}). We report
$\mathit{MS}_{\mathit{strict}} = \text{killed}/\text{live-killable}$ per
operator family and overall.

\smallskip\noindent\textbf{Results.} Table~\ref{tab:ms-kstruct} reports K-Struct mutation
scores per suite and per operator family.

\begin{table*}[t]
\centering
\caption{Live K-Struct mutation score ($\mathit{MS}_{\mathit{strict}}$)
per operator family and per suite, aggregated over the ten benchmarks.}
\label{tab:ms-kstruct}
\small
\setlength{\tabcolsep}{6pt}
\begin{tabular}{@{}lccccr@{}}
\hline
Suite        & F1 deleg.\ & F2 tool & F3 restr.\ & F4 agent
             & $\mathit{MS}_{\mathit{strict}}$ \\
\hline
T-Dev$^{\star}$   & 0.665 & 0.925 & 0.584 & 0.615 & 0.717 \\
T-Rand$^{\star}$  & 0.420 & 0.627 & 0.412 & 0.470 & 0.488 \\
T-E2E$^{\star}$   & 0.527 & 0.667 & 0.462 & 0.590 & 0.560 \\
T-Struct          & \textbf{1.000} & \textbf{1.000} & \textbf{1.000}
                  & \textbf{1.000} & \textbf{1.000} \\
\hline
\end{tabular}
\end{table*}

T-Struct kills every K-Struct-killable mutant by construction (1.000). It
serves as an upper-bound reference suite whose role is to quantify the gap
to the smaller suites, not to claim surprising detection power. The baseline suites fall well short of it.
T-Dev$^{\star}$, the best-performing baseline, reaches 0.717 overall, so even a careful developer smoke suite misses more than one structural fault in four, and T-Rand$^{\star}$ misses over half (0.488). The
per-family scores show where the misses concentrate. T-Dev$^{\star}$ is strong on F1 delegation (0.665) and F2 tool (0.925) because its two-hop
frontier exercises the entry agent's handoffs and the allow edges of the agents it reaches, so perturbing those edges registers as a coordination
change. T-E2E$^{\star}$ is broad but shallow, with one use-tool
obligation per agent, so it catches tool removals but misses the tool
\emph{redirections} that a second obligation would expose (F2 0.667), while T-Rand$^{\star}$, sampling without targeting specific obligations, has lower scores across families. The one
weakness common to every baseline is F3 restriction (0.412 to 0.584). A relaxed or removed restriction changes the coordination trace only when the agent happens to exercise the now-allowed edge, which a positive-polarity suite never explicitly asks it to do, so detecting
restriction faults reliably requires the negative-polarity probes that
only T-Struct carries.

\begin{tcolorbox}[title=RQ1 Summary, breakable]\TRC{The baseline suites miss a substantial
fraction of structural faults even on real agents, with T-E2E$^{\star}$
detecting only $0.560$ overall and T-Rand$^{\star}$ $0.488$, whereas
T-Struct detects all of them. Passing a task benchmark therefore does not
establish that a workflow's coordination structure is correct.}\end{tcolorbox}

\subsection{RQ2: Runtime Criteria K-Outcome and K-Safety}
\label{sec:rq-runtime}

\smallskip\noindent\textbf{Setup.} We apply the two runtime criteria to the same executions used for K-Struct. The killable denominator is the
live-killable count from the inventory: 1{,}010 mutants under K-Outcome
and 177 under K-Safety (a mutant is K-Safety-killable iff some run
fires a tool the original restricts). K-Outcome also evaluates
obligation-preserving operators such as \textsc{Tool-Fault},
\textsc{Ag-Swap}, and \textsc{Ag-PromptDrop}, which can change behavior
without changing the manifest's obligation set.

\smallskip\noindent\textbf{Results.} Table~\ref{tab:ms-runtime} reports the same suites
under all three criteria.

\begin{table}[t]
\centering
\caption{Mutation score per suite under each kill criterion
($\mathit{MS}_{\mathit{strict}}$). The four rows differ only in which
mutants the suite witnesses. The killable denominator is constant across
suites.}
\label{tab:ms-runtime}
\small
\setlength{\tabcolsep}{6pt}
\begin{tabular}{@{}lccc@{}}
\hline
Suite & K-Struct & K-Outcome & K-Safety \\
      & (kbl 983) & (kbl 1{,}010) & (kbl 177) \\
\hline
T-Dev$^{\star}$  & 0.717 & 0.725 & 0.559 \\
T-Rand$^{\star}$ & 0.488 & 0.506 & 0.401 \\
T-E2E$^{\star}$  & 0.560 & 0.563 & 0.503 \\
T-Struct         & \textbf{1.000} & \textbf{1.000} & \textbf{1.000} \\
\hline
\end{tabular}
\end{table}

K-Outcome and K-Safety change the scores differently. K-Outcome
barely moves the scores, lifting T-Dev$^{\star}$ only from 0.717 to
0.725. For the structural operators a change in the coordination trace and a change in task outcome almost always coincide, so K-Outcome is rarely \emph{stronger} than K-Struct. Its contribution is coverage rather than magnitude, since it also
evaluates obligation-preserving operators
(\textsc{Tool-Fault}, \textsc{Ag-Swap}, \textsc{Ag-PromptDrop}), which
have no obligation-set footprint of their own.
K-Safety captures a different failure mode, and here the suites diverge most. The baselines fall to 0.401 to 0.559, below their K-Struct
scores, because catching a forbidden-tool call needs a test that drives
the agent toward the restricted edge, and their obligations only do so occasionally and by chance. An idealized model that never volunteers a
forbidden action would score the baselines 0.000 under K-Safety. In practice live agents do sometimes exercise a relaxed restriction, which
yields some additional kills for the baselines. The remaining gap to T-Struct (1.000) comes from its explicit probe for every restrict edge.

\begin{tcolorbox}[title=RQ2 Summary, breakable]\TRC{The largest gap appears under K-Safety. The baseline suites catch only $0.40$ to $0.56$ of
safety-killable mutants, so a silently relaxed restriction or misrouted tool edge can
pass a task benchmark with no externally visible failure. Only a suite
that explicitly probes the forbidden agent-tool pair, as T-Struct does,
reliably fires the K-Safety trace ($1.000$).}\end{tcolorbox}

\subsection{RQ3: Operator-Level Difficulty}
\label{sec:rq-difficulty}

\smallskip\noindent\textbf{Setup.} Holding the suites and the mutant pool of RQ1 to RQ2
fixed, we disaggregate $\mathit{MS}_{\mathit{strict}}$ to the level of
individual operators (Table~\ref{tab:operators}): one cell per (operator, suite, criterion). We mark a cell \emph{n/a} when its
denominator is empty, that is, when an operator has no live-killable mutant
under that criterion. Under K-Struct and K-Outcome every operator has
live-killable mutants, but under K-Safety only \textsc{Res-Relax} and
\textsc{Tool-Red} do, since only they make a restricted tool callable. The
remaining operators cannot produce a forbidden call, so their K-Safety
cells are \emph{n/a}. (This run-level \emph{n/a} is distinct from the
manifest-only \emph{n/a} of Section~\ref{sec:inventory},
which reflects an operator a non-executing oracle cannot score at all.) We
then check whether the per-operator difficulty ranking is stable across
kill criteria. 

\begin{figure*}[t]
\centering
\includegraphics[width=0.9\textwidth]{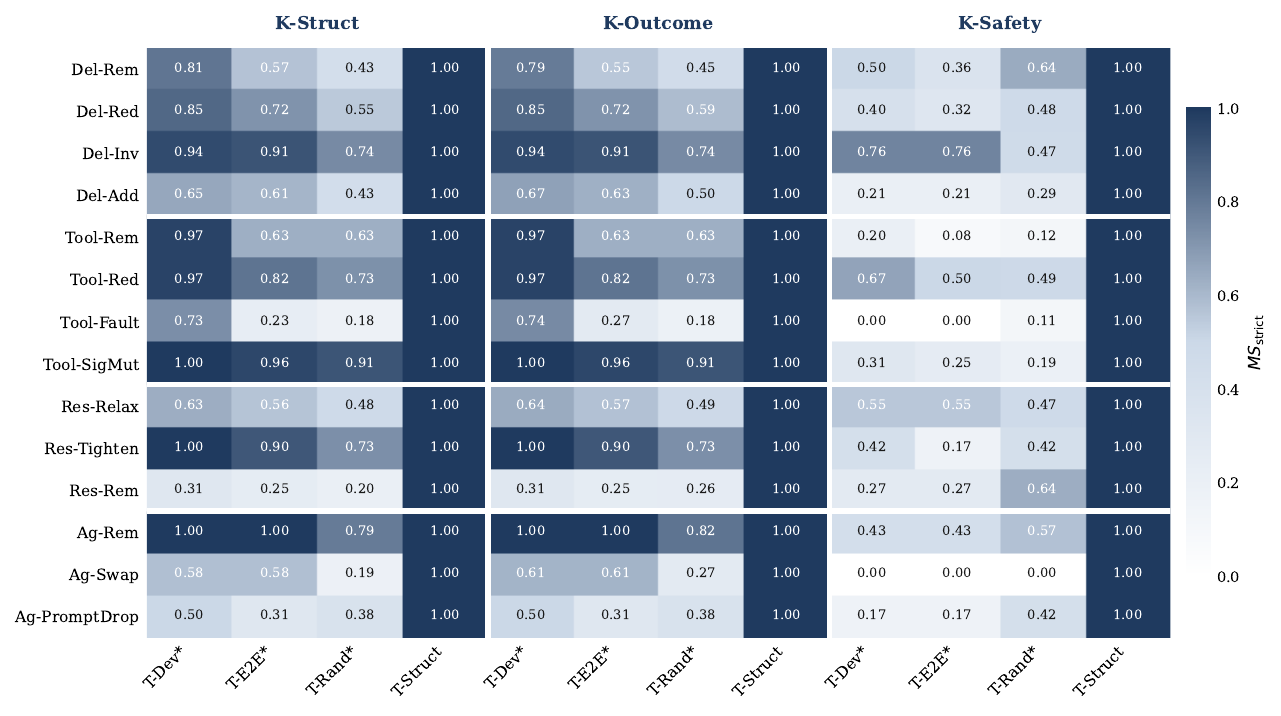}
\caption{$\mathit{MS}_{\mathit{strict}}$ across all 14~operators
(rows, grouped by family F1 to F4), four suites, and three kill criteria.  Vertical bands separate criteria. Horizontal bands separate operator families. Every operator has live-killable mutants under K-Struct and K-Outcome. Under K-Safety only \textsc{Res-Relax} and \textsc{Tool-Red} do, so the other operators' K-Safety cells are \emph{n/a} (gray).}
\label{fig:heatmap-msstrict}
\end{figure*}

\smallskip\noindent\textbf{Results.}
Figure~\ref{fig:heatmap-msstrict} summarizes the full 
$\mathit{MS}_{\mathit{strict}}$ matrix, and
Table~\ref{tab:ms-by-operator-koutcome} gives the K-Outcome column numerically. The difficulty ranking is stable across K-Struct and K-Outcome, under which every operator yields live-killable mutants. Under K-Safety only \textsc{Res-Relax} and \textsc{Tool-Red} have live-killable mutants, since only they make a restricted tool callable, so the remaining operators' K-Safety cells are \emph{n/a}.

\begin{table}[t]
\centering
\caption{Per-operator $\mathit{MS}_{\mathit{strict}}$ under
K-Outcome, aggregated over the ten benchmarks.}
\label{tab:ms-by-operator-koutcome}
\small
\setlength{\tabcolsep}{4pt}
\begin{tabular}{@{}lcccc@{}}
\hline
Operator & T-Dev$^{\star}$ & T-Rand$^{\star}$ & T-E2E$^{\star}$ & T-Struct \\
\hline
\textsc{Del-Rem}        & 0.769 & 0.564 & 0.564 & 1.000 \\
\textsc{Del-Red}        & 0.684 & 0.389 & 0.474 & 1.000 \\
\textsc{Del-Inv}        & 0.857 & 0.657 & 0.771 & 1.000 \\
\textsc{Del-Add}        & 0.554 & 0.500 & 0.491 & 1.000 \\
\textsc{Tool-Rem}       & 1.000 & 0.710 & 0.855 & 1.000 \\
\textsc{Tool-Red}       & 0.961 & 0.682 & 0.649 & 1.000 \\
\textsc{Tool-Fault}     & 0.726 & 0.113 & 0.258 & 1.000 \\
\textsc{Tool-SigMut}    & 1.000 & 0.867 & 0.956 & 1.000 \\
\textsc{Res-Relax}      & 0.616 & 0.488 & 0.568 & 1.000 \\
\textsc{Res-Tighten}    & 0.968 & 0.619 & 0.651 & 1.000 \\
\textsc{Res-Rem}        & 0.280 & 0.190 & 0.200 & 1.000 \\
\textsc{Ag-Rem}         & 1.000 & 0.789 & 1.000 & 1.000 \\
\textsc{Ag-Swap}        & 0.628 & 0.442 & 0.605 & 1.000 \\
\textsc{Ag-PromptDrop}  & 0.378 & 0.270 & 0.324 & 1.000 \\
\hline
\end{tabular}
\end{table}

Restriction and tool faults fail for different reasons: restriction faults
are silent, whereas tool faults require the suite to actually hit the
affected tool.
Subtlety dominates the restriction family. \textsc{Res-Rem} is the least detected
operator (T-Dev$^{\star}$ 0.280, T-E2E$^{\star}$ 0.200, T-Rand$^{\star}$
0.190 vs T-Struct 1.000), because silently removing a restriction adds capability without causing an immediate failure, and an agent seldom
volunteers a forbidden action even once it is allowed.
\textsc{Res-Relax} is somewhat easier (0.49 to 0.64) and
\textsc{Res-Tighten} easier still (T-Dev$^{\star}$ 1.000), since
tightening removes an allowed capability the smoke tests already
exercise.

Budget dominates the tool family. These operators are easy for
T-Dev$^{\star}$ (0.74 to 1.00), which exercises every allow edge on the
entry frontier, but hard for the budget-limited suites, with T-E2E$^{\star}$'s single use-tool obligation per agent leaving it at
0.258 on \textsc{Tool-Fault} and T-Rand$^{\star}$ dipping to 0.113
because one prompt rarely drives the agent through the faulting tool. The same pattern appears under both K-Struct and K-Outcome
(Figure~\ref{fig:heatmap-msstrict}), suggesting that the difficulty comes
mainly from the fault type rather than from the chosen criterion.

Finally, \textsc{Ag-Swap} and \textsc{Ag-PromptDrop}, which perturb only instruction text and are invisible to a manifest-only oracle, are
detected here. T-Struct kills 1.000 of both, and even T-Dev$^{\star}$ reaches 0.628 (\textsc{Ag-Swap}) and 0.378
(\textsc{Ag-PromptDrop}), because an agent routes differently when an
instruction is swapped or dropped and the structural suite witnesses the
change.

\begin{tcolorbox}[title=RQ3 Summary, breakable]\TRC{Restriction-removal
(\textsc{Res-Rem}) is the least detected fault for the baselines ($0.19$ to $0.28$ vs $1.000$). It broadens capability rather than breaking
functionality, so an agent only occasionally exercises the newly allowed edge. Execution through the SDK also allows the obligation-preserving operators
to be evaluated: \textsc{Ag-Swap} and \textsc{Ag-PromptDrop} can change which specialist the workflow routes to, and \textsc{Tool-Fault} can
change how an agent recovers, none of which a manifest comparison would
reveal.}\end{tcolorbox}

\subsection{Running Example: \texttt{oai\_customer\_service}}
\label{sec:worked-example}

We illustrate the criteria on the running example.
A \textsc{Del-Rem} mutant that severs
\texttt{triage\_agent}\,$\to$\,\texttt{faq\_agent} makes the triage agent report that no \texttt{faq\_agent} subagent is available for handoff. Every
suite whose obligation set contains that delegation kills it under
K-Struct and K-Outcome. A \textsc{Tool-Rem} on
\texttt{(faq\_agent, faq\_lookup\_tool)} yields a refusal in place of the
original lookup, again a structural and outcome kill. The restriction
faults are where the suites separate. Under \textsc{Res-Relax} on
\texttt{(faq\_agent, update\_seat)} the original agent reports the tool forbidden, but the mutant, now allowed, only \emph{sometimes} invokes
\texttt{update\_seat} of its own accord. T-Struct, which carries an
explicit negative-polarity probe for that edge, drives the violation and fires K-Safety, whereas the baseline suites observe the leak only on the
runs where the agent volunteers the call. Finally, an \textsc{Ag-PromptDrop} that clears \texttt{faq\_agent}'s instruction hint, invisible to a manifest-only oracle, changes how the agent handles the request, a
divergence in the verdict or step sequence that T-Struct catches under
K-Outcome. (\textsc{Ag-Swap} is an identity on this workflow, since the two specialists share a prompt hint, so its routing effect appears only in workflows whose agents carry distinct instructions.) These per-mutant behaviors
underlie the aggregate scores in
Tables~\ref{tab:ms-kstruct} and \ref{tab:ms-by-operator-koutcome}.

\subsection{Threats to Validity}
\label{sec:threats}

\smallskip\noindent\textbf{Non-determinism.}
Every kill is decided by stochastic LLM execution, so an individual kill
can be flaky. We mitigate this with $k{=}3$ repeats and a per-test
majority vote, and the residual variance does not affect the qualitative
ranking T-Struct $\gg$ T-Dev$^{\star}$ $>$ T-E2E$^{\star}$ $>$
T-Rand$^{\star}$. A smaller pilot sweep over a subset of the mutants
reproduced the same qualitative outcome, with T-Struct above all three
baselines and T-Rand$^{\star}$ weakest, so this ordering is not an
artifact of run-to-run variance.

\smallskip\noindent\textbf{Test selection by structural incidence.}
To bound cost, each mutant is executed only on the obligations structurally incident to its changed element rather than on the full
obligation set. This is a manifest-level static computation, not a
behavioral simulation, but it is an approximation. If a mutant altered behavior on a non-incident obligation, we would undercount its kills.
Because our operators are first-order and edit a single structural element, incident obligations are where divergence is expected. The
approximation trades recall for tractability.

\smallskip\noindent\textbf{Capping and sampling.}
We cap each operator at 50 mutants per benchmark, and at 10 for \texttt{octagon\_vc\_agents} (167 obligations) to keep the sweep
tractable. Octagon therefore contributes fewer mutants to the pooled denominators than it would at the full cap. The seed
(\texttt{20260605}) fixes mutant sampling and suite construction but not
the model. A larger compute budget would let every operator reach the
full 50/op on every benchmark.

\smallskip\noindent\textbf{Framework and model scope.}
All ten benchmarks are OpenAI Agents SDK-derived workflows, executed on
\texttt{gpt-4.1-mini}. The operator catalog and the
K-Struct/K-Outcome/K-Safety triple transfer to any framework that exposes explicit agent, tool, allow, restrict, and delegate primitives
(notably LangGraph and AutoGen), but the runtime adapter must be re-implemented per framework. We hypothesize that the relative trends will
carry over, but this remains to be tested, and absolute scores will vary
with the underlying model.

\section{Related Work}
\label{sec:related-work}

This work draws on classical mutation testing, LLM/agent testing, and runtime fault injection.

Mutation analysis was introduced by DeMillo et
al.~\cite{demilli1978hints} and surveyed by Jia and
Harman~\cite{jia2011analysis} and Papadakis et
al.~\cite{papadakis2019mutation}. Its standard pipeline, small syntactic
mutants, kill criteria based on test outcomes, and equivalent-mutant
filtering, has been instantiated for Java (MuJava~\cite{ma2005mujava},
PIT~\cite{coles2016pit}), C (Milu~\cite{jia2008milu}), JavaScript, and
SQL, and Just et al.~\cite{just2014mutants} showed that the mutation score is a useful proxy for real-fault detection, a finding corroborated
by Andrews et al.~\cite{andrews2005mutation} against manually seeded
faults. Higher-order, selective, and similarity-based operators reduce its cost~\cite{offutt1992investigations,jia2009higher,zhang2014predictive}.
We restrict ourselves to first-order mutants because the structural
operator catalog is small and agent runtimes are expensive to execute, and our equivalent-mutant audit follows the manual-rating protocol of
Just et al.~\cite{just2014mutants}. We carry this machinery to a new target, the coordination structure of a multi-agent workflow, mutated at
the specification level rather than in program source code.

That target is largely unaddressed by existing agent evaluation.
End-to-end benchmarks such as AgentBench~\cite{liu2023agentbench},
GAIA~\cite{mialon2023gaia}, API-Bank~\cite{li2023apibank},
ToolLLM~\cite{qin2024toolllm}, SWE-bench~\cite{jimenez2024swebench}, and
WebArena~\cite{zhou2024webarena} measure task success on fixed tasks but
say nothing about the fault-detection power of the suite itself.
Structural coverage for multi-agent
workflows~\cite{kahani2025structuralagentcoverage} measures whether
declared agents, tools, and edges are exercised. Coverage and mutation
score are complementary, and our experiments confirm that high coverage does not imply a high mutation score for non-structural suites.
Capability-based NLP testing frameworks such as
CheckList~\cite{ribeiro2020checklist} and
TextFlint~\cite{wang2021textflint} ask a behavioral question
complementary to ours (does the suite detect faults?), and our K-Outcome
criterion can in principle wrap any such task-success oracle.

Input-perturbation work differs from ours in target, it perturbs the
model's input rather than the workflow specification. Adversarial probing of LLMs, including
TextAttack~\cite{morris2020textattack}, prompt-injection
benchmarks~\cite{greshake2023prompt,perez2022red,zou2023universal}, and
jailbreak studies~\cite{wei2024jailbroken}, manipulates inputs to elicit
unsafe outputs, whereas we perturb the system \emph{specification} to evaluate test suites. The two address different threat models. Our
\textsc{Res-Relax} operator is the closest in spirit, since it asks
whether a suite would catch a silent policy relaxation introduced by a
developer rather than an attacker.

Two testing traditions are closest to our operators. Mutation testing has been adapted to
neural networks (DeepMutation~\cite{ma2018deepmutation},
MuNN~\cite{shen2018munn}) by perturbing weights, activations, or training
data. Those operators target the learned model, whereas ours target the deterministic workflow code surrounding it. Metamorphic testing for
ML~\cite{xie2011testing,zhang2018deeproad} exploits relations between
inputs and outputs, and our K-Struct criterion is in the same spirit, 
treating $\mathsf{obs}^{\mathit{str}}_S(x) \neq \mathsf{obs}^{\mathit{str}}_{S'}(x)$ as a structural metamorphic relation over a specification perturbation rather than an
input transformation. Finally, runtime fault-injection frameworks such as Chaos Monkey~\cite{basiri2016chaos}, Jepsen~\cite{kingsbury2014jepsen},
and Fault Injection in Production~\cite{alvaro2015lineage} perturb a
running distributed system to test resilience. Our \textsc{Tool-Fault} operator is the direct analog at the agent layer, where a tool stub
returns an error or a malformed payload and the test asks whether the
surrounding agent recovers.

\section{Conclusion}
\label{sec:discussion}

The main result of this paper is that task-success-style tests miss many workflow-level faults, especially permission relaxations that do not break the final answer. In our ten-workflow study, the strongest baseline
reaches $0.717$ under K-Struct but only $0.559$ under K-Safety, while the full-obligation T-Struct suite serves as the reference upper bound. We
obtained this result with mutation analysis for multi-agent workflow
specifications, fourteen structural operators in four families, three behavioral kill criteria (K-Struct, K-Outcome, K-Safety) decided by
execution through the OpenAI Agents SDK, and an evaluation of $1{,}084$
mutants over $29{,}601$ agent executions. Executing
against real agents, rather than a manifest-only oracle, also makes
the obligation-preserving operators (\textsc{Tool-Fault}, \textsc{Ag-Swap},
\textsc{Ag-PromptDrop}) scorable. Across the ten OpenAI Agents SDK workflows we study, task-success evaluation does not reliably detect coordination-level faults, while the full-obligation suite exposes those faults; whether the same trends hold on other frameworks and models remains to be tested.

The next step is to make the generator mutation-aware. It could regenerate prompts that target surviving mutants, producing tests that
explicitly differentiate the original from each mutant. Selective
higher-order operators~\cite{jia2009higher} would capture the multi-change regressions that first-order mutants miss. Porting the
runtime adapter to LangGraph, AutoGen, and Semantic Kernel would broaden the empirical scope, and repeating the sweep across models would test how
far the detection patterns we report are model-independent.

\bibliographystyle{unsrt}
\bibliography{references}

\end{document}